\documentclass[manuscript]{aastex631}

\usepackage{bm}

\begin{document}
\title{Observational Comparison Between Confined and Eruptive Flares: Magnetohydrodynamics Instability Parameters in a Similar Magnetic Configuration}

\author{Kouhei Teraoka}
\affiliation{Department of Earth and Planetary Science, The University of Tokyo, 7-3-1 Hongo, Bunkyo-ku, Tokyo 113-0033, Japan;
\url{k-teraoka892@g.ecc.u-tokyo.ac.jp}}
\affiliation{Institute of Space and Astronautical Science, Japan Aerospace Exploration Agency, 3-1-1 Yoshinodai, Chuo, Sagamihara, Kanagawa 252-5210, Japan;
\url{shimizu.toshifumi@jaxa.jp}}

\author{Daiki Yamasaki}
\affiliation{Institute of Space and Astronautical Science, Japan Aerospace Exploration Agency, 3-1-1 Yoshinodai, Chuo, Sagamihara, Kanagawa 252-5210, Japan;
\url{shimizu.toshifumi@jaxa.jp}}

\author{Yusuke Kawabata}
\affiliation{National Astronomical Observatory of Japan, 2-21-1 Osawa, Mitaka, Tokyo 181-8588, Japan}

\author{Shinsuke Imada}
\affiliation{Department of Earth and Planetary Science, The University of Tokyo, 7-3-1 Hongo, Bunkyo-ku, Tokyo 113-0033, Japan;
\url{k-teraoka892@g.ecc.u-tokyo.ac.jp}}

\author{Toshifumi Shimizu}
\affiliation{Department of Earth and Planetary Science, The University of Tokyo, 7-3-1 Hongo, Bunkyo-ku, Tokyo 113-0033, Japan;
\url{k-teraoka892@g.ecc.u-tokyo.ac.jp}}
\affiliation{Institute of Space and Astronautical Science, Japan Aerospace Exploration Agency, 3-1-1 Yoshinodai, Chuo, Sagamihara, Kanagawa 252-5210, Japan;
\url{shimizu.toshifumi@jaxa.jp}}

\begin{abstract}
Unstable states of the solar coronal magnetic field structure result in various flare behaviors. In this study, we compared the confined and eruptive flares that occurred under similar magnetic circumstances in the active region 12673, on 2017 September 6, using the twist number, decay index, and height of magnetic field lines to identify observational behaviors of the flare eruption. We investigated the parameters from the magnetic field lines involved in an initial energy release, which were identified from the positions of the core of flare ribbons, i.e., flare kernels. The magnetic field lines were derived by nonlinear force-free field modeling calculated from the photospheric vector magnetic field obtained by the \textit{Solar Dynamics Observatory} \textit{SDO}/Helioseismic and Magnetic Imager, and flare kernels were identified from the 1600 \AA\ data obtained by the \textit{SDO}/Atmospheric Imaging Assembly. The twist number of all the magnetic field lines in the confined flare was below 0.6; however, the twist number in seven out of twenty-four magnetic field lines in the eruptive flare was greater than 0.6. These lines were tall. It is found that the decay index is not a clear discriminator of the confined and eruptive flares. Our study suggests that some magnetic field lines in the kink instability state may be important for eruptive flares, and that taller magnetic field lines may promote flare eruption.
\end{abstract}


\section{Introduction} \label{sec:intro}
Solar flares are sudden releases of magnetic energy stored in magnetic structures formed in the solar atmosphere. Numerous numbers of theoretical models have been proposed to date; however, the CSHKP model \citep{1964NASSP..50..451C, 1966Natur.211..695S, 1974SoPh...34..323H, 1976SoPh...50...85K} is widely used to describe the temporal evolution of magnetic structures in good agreement with observational behaviors, such as temporal evolution of cusp structure in soft X-ray \citep{1992PASJ...44L..63T}, hard X-ray sources in impulsive phase \citep{1994Natur.371..495M}, flare ribbons in H$\alpha$ \citep{2003ApJ...586..624A}, and eruptions of dark filaments \citep{1988ApJ...324.1132M}. A bundle of highly twisted magnetic field lines called magnetic flux ropes (MFRs) may develop along the polarity inversion line (PIL) in the corona. Pre-existing magnetic field lines connect two opposite magnetic islands on both sides of the PIL and mostly overlap with the highly sheared MFR. When the MFR is in an unstable magnetohydrodynamics (MHD) state and begins to move upward, it may drive the overlying magnetic fields and trigger magnetic reconnection \citep{1958IAUS....6..123S, 1957JGR....62..509P, 1964NASSP..50..425P}, which converts magnetic energy into thermal and kinetic energy. After magnetic reconnection, flare ribbons appear in the chromosphere. The magnetic field lines at the core of flare ribbons as observed as flare kernels connect to the reconnected magnetic field lines, and transport the heated and accelerated particles generated by magnetic reconnection. These particles form flare ribbons upon heating the surrounding plasma in the chromosphere.
The magnetic fields at the upper side of magnetic reconnection may move upward because of the MHD instabilities, driving the upper overlapping magnetic field lines. If this upward motion continues, some of the magnetic field lines and plasma may escape into space. This escape event is considered a coronal mass ejection (CME). 
Flares that produce large CMEs are called eruptive flares, whereas those that produce no CMEs or only small CMEs may be called confined flares. The latter may be also called failed eruption flares \citep{2003ApJ...595L.135J}.
It remains unclear how coronal magnetic structures become unstable before flares occur and why the sequence of flares may result in either eruption or confinement.

Theoretically, at least two types of MHD instability—kink instability \citep{1979SoPh...64..303H, 2005ApJ...630L..97T} and torus instability \citep{1964PhFl....7..278Y, 2006PhRvL..96y5002K}—could be responsible for the shift toward unstable conditions in coronal magnetic structures and driving the dynamic evolution of large-scale magnetic structures. This could result in a variety of behaviors during the flare onset evolution. Kink instability occurs when highly twisted magnetic field lines are converted into a writhe. The twist number, which represents the number of turns in the twisted magnetic field lines, is defined as
\begin{equation}
   T_{w} = \frac{1}{4\rm{\pi}}\int_{l}^{ } \frac{\bm{\nabla\times B\cdot B}}{B^{2}} dl,
\end{equation}
where $l$ is the magnetic field line, $dl$ is an element of length along magnetic field line, $l$, and $\bm{B}$ is the magnetic flux density. The twist number is also a proxy for the magnetic free energy. 

Torus instability refers to the instability in which torus-shaped and twisted magnetic field lines are unstable against expansion. This unstable state is caused by a rapid decrease in the external poloidal magnetic flux density. Because the dominant force in the corona is the Lorentz force, upward hoop and downward strapping forces act on the apex of the toroidal magnetic field lines. The decay index, which represents the decay speed with the height of the external poloidal magnetic flux density, is defined as 
\begin{equation}
    n = -\frac{z}{|\bm{B_{ex}}|}\frac{\partial |\bm{B_{ex}}|}{\partial z},
\end{equation}
where $\bm{B_{ex}}$ is the external poloidal magnetic flux density and z is the apex height of the toroidal magnetic field lines. This definition implies that $|\bm{B_{ex}}| \propto z^{-n}$. The decay index is extensively used to analyze the coronal magnetic structure in the center of the flaring region to determine whether an active region (AR) has the potential to produce eruptive flares. Kink and torus instabilities in flares have been investigated by many researchers—for example, \citet{2008ApJ...679L.151L}, \citet{2010ApJ...725L..38G}, \citet{2011ApJ...732...87C}, \citet{2012ApJ...748L...6N}, \citet{2012ApJ...760...17I}, \citet{2013ApJ...778L..36L, 2016ApJ...818..148L}, and \citet{2016ApJ...831L..18L}—however, the conditions for flare eruptivity have not yet been accurately defined.

In addition to surveying each instability, a combined study of the two instabilities was conducted to investigate the conditions for distinguishing between confined and eruptive flares. Kink and torus instabilities were investigated in a laboratory plasma experiment, in which a torus-like twisted flux rope is tied down to the base plate \citep{2015Natur.528..526M}. Additionally, by changing the edge safety factor (inverse of the twist number; $q_{a} = 1/T_{w}$) and decay index, it can be statistically determine whether a MFR eruption occurs. Four distinct parameter regimes were identified in the safety factor—decay index diagram: eruptive (kink-unstable and torus-unstable), stable (kink-stable and torus-stable), failed kink (torus-stable but kink-unstable), and failed torus (kink-stable but torus-unstable). Kink instability appears below $q_{a} \approx 0.8$ (the twist number is greater than $T_{w} \approx 1.25$). Torus instability appears when the decay index is greater than $n \approx 0.8$. Most MFRs erupted in the eruptive regime. 
\citet{2018ApJ...864..138J} compared the observations of eruptive flares with those of confined flares to determine whether the conclusions of \citet{2015Natur.528..526M} can be applied to actual solar flare cases. The confined and eruptive flares were not clearly distinguished in terms of the twist number. The 12 flares with $n \gtrsim 0.8$ are all eruptive; however, confined flares are mixed with eruptive flares in the regimes $n \lesssim 0.8$, twelve out of the twenty-six flares with $n \lesssim 0.8$ are confined, and the remaining 14 flares are eruptive. Thus, $n \simeq 0.8$ is not a necessary condition for eruptive flares. Their results do not show the presence of the four regimes derived by \citet{2015Natur.528..526M}. A reason why the observations are not shown, as shown in laboratory experiments, is that \citet{2015Natur.528..526M} used similar MFRs in a controlled laboratory environment, whereas \citet{2018ApJ...864..138J} chose a variety of MFRs in their observations. Therefore, confined and eruptive flares produced with similar MFR configurations should be compared for observational purposes. As examples, the two on-disk X-class flares produced within 3 h on 2017 September 6, in AR 12673. The former applies to the X2.2 confined flare starting at 08:57 UT, and the latter to the X9.3 eruptive flare starting at 11:53 UT. Because they occurred within only 3 h of each other and at the same location, the magnetic field configurations are similar.
The remainder of this paper is organized as follows. Section \ref{sec:obs} describes the observations, and Section \ref{sec:data} describes the data analysis. In Section \ref{sec:results}, we present the results, followed by a discussion in Section \ref{sec:discussion}. Finally, we summarize the conclusions drawn in Section \ref{sec:conclusions}.

\section{Observations} \label{sec:obs}
\subsection{SDO Observations} \label{subsec:sdo}
The Helioseismic and Magnetic Imager (HMI) \citep{2012SoPh..275..207S,2012SoPh..275..229S} 
on board the \textit{Solar Dynamics Observatory} (\textit{SDO}) \citep{2012SoPh..275....3P}
provides full-disk photospheric 3D magnetic field vector magnetograms with 4096 $\times$
4096 pixels, each with a pixel size of 0.5 arcseconds. The HMI samples six points spanning in the range of
$\pm$ 172.5 m\AA\ around the FeI 6173.3 \AA\ and measures six polarizations of $I \pm Q$, $I \pm U$, and $I \pm V$. The HMI generated 36 full-disk filtergrams every 135 s, from which the vector magnetic field is determined. We used the Space-weather HMI Active Region Patches (SHARPs) \citep{2014SoPh..289.3549B} data series (\textbf{hmi.sharp\_720s}). The data 
have three components, that is, field strength (Mx cm$^{-2}$), inclination angle (degrees), and 
azimuth angle (degrees) (line of sight coordination), and are automatically divided into each active region \citep{2014SoPh..289.3483H}. A tapered temporal average is performed every 720 s using 360 filtergrams collected over a 
1350 s interval to produce 36 corrected, filtered, and co-registered images to reduce noise and 
minimize the effects of solar oscillations \citep{2012SoPh..275..285C}. We used the data as the bottom condition to model the magnetic field lines in the corona (Section \ref{subsec:nlfff}).

The Atmospheric Imaging Assembly (AIA) \citep{2012SoPh..275...17L} onboard the \textit{SDO} provides full-disk EUV and UV images of the solar atmosphere with 4096 $\times$ 4096 pixels, each with a pixel size of 0.6 arcseconds. We used 1600 \AA\ data (see the chromosphere) with a cadence of 24 s to identify the location of the flare kernels. We obtained a Level 1 data series (\textbf{aia.lev1\_uv\_24s}).

\subsection{Target Active Region \label{subsec:targetar}}
\begin{figure*}[ht!]
\centering
\includegraphics[scale=0.9,clip]{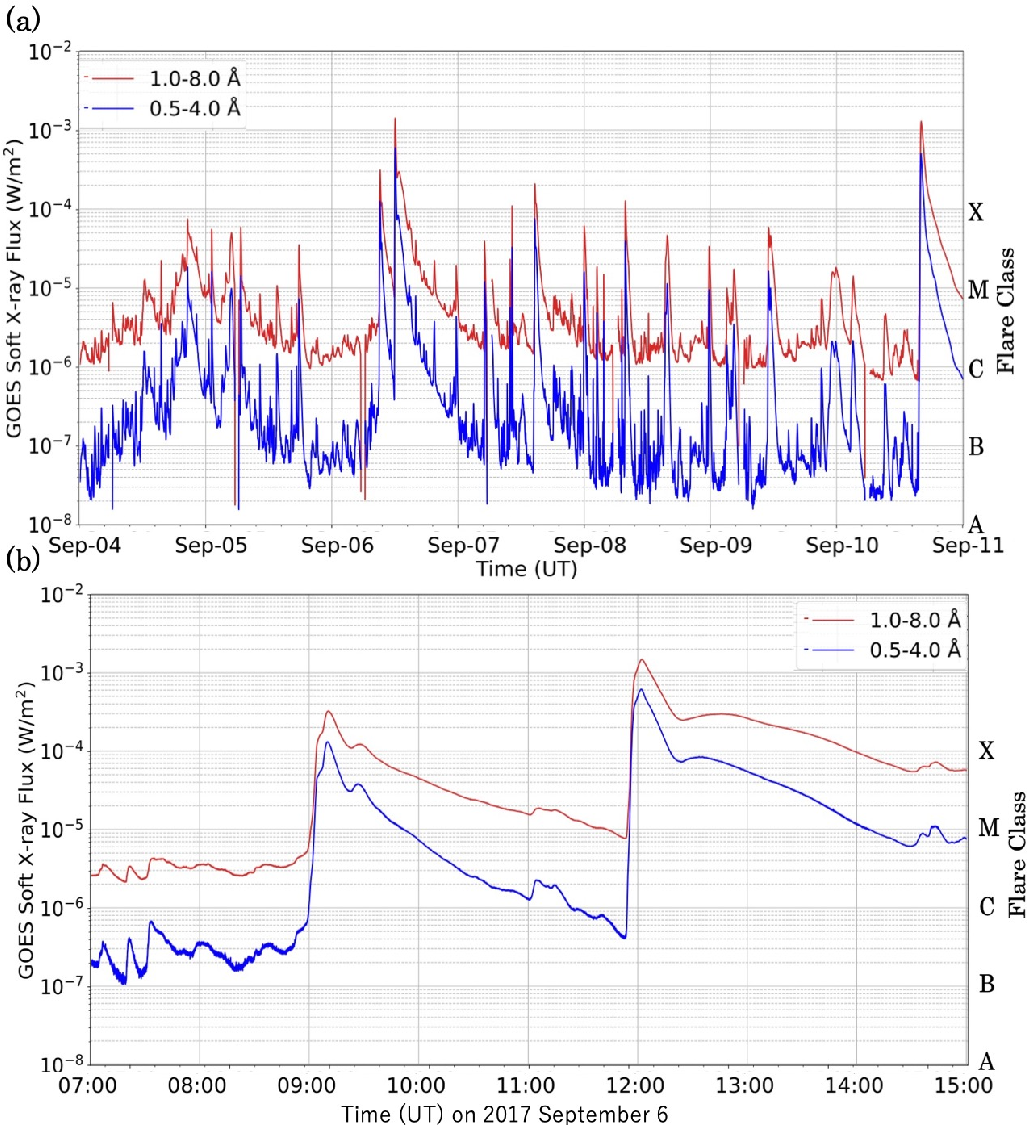}
\caption{GOES soft X-ray flux (a) from 00:00 UT on 2017 September 4, to 00:00 UT on 2017 September 11, and (b) a zoom up including the two flares from 07:00 UT to 15:00 UT on September 6. The vertical axis represents the GOES soft X-ray flux ($\mathrm{W/m^{2}}$) in 1–8 \text{\AA} by red and in 0.5–4 \text{\AA} by blue.
\label{fig:100}}
\end{figure*}

AR 12673 appeared on the solar disk on 2017 August 31, and was rotated behind the limb by 2017 September 10. This AR was the $\beta \delta \gamma$ type and had four X-class and 27 M-class flares from September 4 to 10 (as shown in Figure \ref{fig:100} (a)). 
In particular, the X9.3 flare is the largest flare in solar cycle 24. From September 4–10, many M-class flares occurred, but no M-class flares were observed half a day prior to the onset of the X2.2 flare on September 6. Most flares lasted for a few tens of minutes on soft X-rays, except for the M1.1 flare on September 9 (157 min). 
According to \citet{2017ApJL...849...L21}, \citet{2018A&A...619..A100}, and \citet{2020ApJ...894...29B} 
, AR 12673 evolved dramatically with strong flux emergence and shearing. \citet{2021ApJ...908..132Y} speculated that AR 12673 had three MFRs on September 4, and the M5.5 flare may change the coronal magnetic structure and suppressed two MFRs. One of them may trigger the succeeding X2.2 confined and X9.3 eruptive flares.

\subsection{Target Flares \label{subsec:target:Targetf}}
On 2017 September 6, AR 12673 was located at S09W42. Table \ref{tab:deluxesplit} summarizes the fundamental properties of these flares.
\begin{table*}[ht!]
\centering
\caption{Fundamental characteristics of the confined and eruptive flares in AR 12673 \label{tab:deluxesplit}}
\begin{tabular}{lcc}
\hline \hline
{} & Confined Flare & Eruptive Flare \\
\tableline
Flare Duration & 08:57 UT - 09:17 UT & 11:53 UT - 12:10 UT \\

Flare Class & X2.2 & X9.3 \\

CME Onset time & 09:48 UT & 12:24 UT \\
{} & (at 2.86 solar radius) & (at 3.93 solar radius) \\

CME Linear Speed & 391 km/s & 1571 km/s \\

CME Angular Width & $80^{\circ}$ & $360^{\circ}$ (halo CME) \\

CME Mass & $3.3 \times 10^{15}$ g & $2.9 \times 10^{16}$ g \\

CME Kinetic Energy & $2.5 \times 10^{30}$ erg & $3.6 \times 10^{32}$ erg \\
\tableline
\end{tabular}
\begin{flushleft}
\tablecomments{CME onset time and position were determined from the first detection data of the \textit{SOHO}/LASCO C2 coronagraph. The numbers in this table were obtained from the LASCO CME catalog (\url{https://cdaw.gsfc.nasa.gov/CME_list/}).}
\end{flushleft}
\end{table*}
The duration of both the flares was approximately 20 min. The soft X-ray flux in the eruptive flare is 4.2 times larger than that in the confined flare. The CME in the confined flare occurred 51 min after the flare onset, and that in the eruptive flare occurred 31 minutes after the flare onset. The linear speed of the CME in the eruptive flare was four times faster than that in the confined flare. The CME angular width in the confined flare was $80^{\circ}$ and that in the eruptive flare was $360^{\circ}$ (halo CME). The CME mass in the eruptive flare was 8.8 times larger than that in the confined flare. The CME kinetic energy in the eruptive flare was 144 times greater than that in the confined flare. Thus, the CME kinetic energy per flare thermal energy in the eruptive flare was 34 times higher. This means that compared to the flare scale, the CME energy was much larger in the eruptive flare than in the confined flare. 
Although the LASCO CME catalog does record a tiny CME after the X2.2 flare, it was so faint compared with the CME associated with the X9.3 flare that we could call the flare as a confined flare.
\citet{2018ApJ...867L...5L} and \citet{2019ApJ...870...97Z, 2020ApJ...890...10Z} also regard the X2.2 flare as a confined flare.

Figure \ref{fig:100} (b) shows the temporal evolution of the GOES soft X-ray flux, including the confined and eruptive flares. 
At first glance, the time evolutions of the two flares were similar. In the confined flare, the soft X-ray flux began to increase sharply at 09:00 UT, and reached its first peak at 09:04 UT. After a few minutes of gradual increase, it reached its second peak at 09:10 UT, followed by a gradual decay. A subtle increase was obtained at 09:28 UT. During the eruptive flare, the soft X-ray flux sharply increased at 11:54 UT and reached its first peak at 11:58 UT. After a few minutes of gradual increase, it reached a second peak at 12:02 UT, followed by a gradual decay phase until 12:22 UT, after which the light curve flattened.

\begin{figure*}[ht!]
\centering
\includegraphics[scale=0.94,clip]{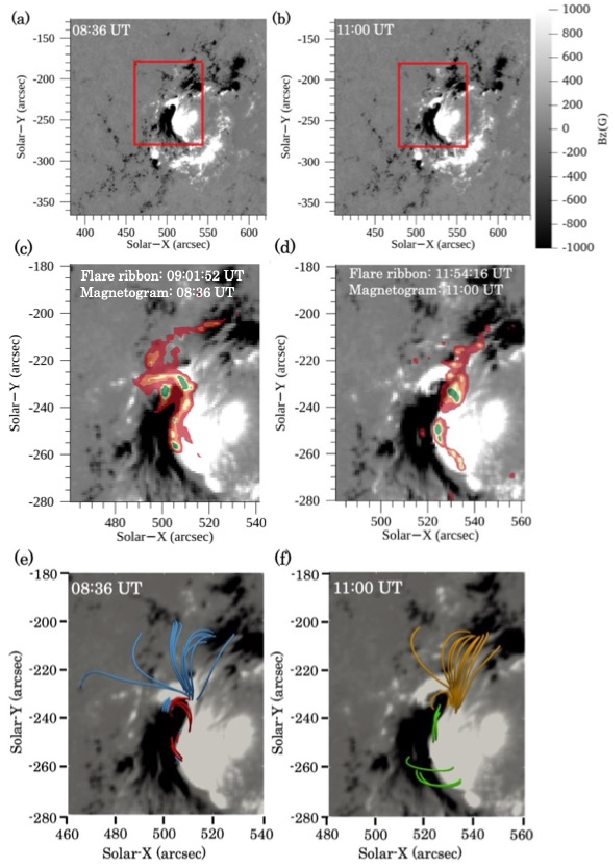}
\caption{Panels (a), and (b): Preprocessed (see Section \ref{subsec:nlfff}) magnetogram (a) before the confined flare, and (b) before the eruptive flare. The X-axis is solar-X (arcsecond) and the Y-axis is solar-Y (arcsecond) at the heliocentric coordinate. The color bar shows the flux density ranging from $-$1000 G (black) to 1000 G (white).  Panels (c), and (d): Pixel location of the flare kernels (green areas) for the confined flare and eruptive flare respectively, identified in the first frame of the AIA 1600 \AA\ measurements. Green areas are defined as over 2000 DN/s/pixel, yellow areas as over 1000 DN/s/pixel, orange areas as over 500 DN/s/pixel, and red areas as over 200 DN/s/pixel. Panels (e), and (f): Magnetic field lines involved in the initial energy release rooted in the flare kernels (pixels of the green areas in (c) and (d)). The blue and red lines in (e) for the confined flare, whereas the orange and green lines in (f) for the eruptive flare. The field of view in (c), (d), (e), and (f) is shown by red rectangle in (a) and (b).
\label{fig:101}}
\end{figure*}

To obtain the boundary data for the 3D magnetic field numerical calculations (explained in Section \ref{subsec:nlfff}), we chose the SHARP data at 08:36 UT (for the confined flare) and 11:00 UT (for the eruptive flare) (Figures \ref{fig:101} (a) and (b)).
Because the boundary data should fulfill the force-free condition, we selected those for which the observed time was sufficient before the flare onset.
Note that we could not obtain the SHARP data before 08:36 UT due to an Earth eclipse. To determine the position of the flare kernels, we chose the AIA 1600 \AA\ data at 09:01:52 UT (for the confined flare) and 11:54:16 UT (for the eruptive flare) (the positions of the flare kernels are the green areas in Figures \ref{fig:101} (c) and (d)).
The AIA data were binned from a pixel size of 0.6 arcseconds to 1.0 arcsecond and the local SHARP data were co-aligned into the global AIA data using the \textbf{index2map} and \textbf{plot\_map} routine available through the Solar-SoftWare (SSW) package \citep{1998SoPh..182..497F}
(See Section \ref{subsec:ribbon} for the accuracy of the co-alignment). Data were collected from the flares during the onset time, and the first flame was selected when the flare kernels initially appeared. We did not use the flare peak data because we wanted to know the flare onset process.
Many authors, i.e., \citet{2017ApJL...849...L21}, and \citet{2018A&A...619..A100}, have already investigated the flares in the other AIA channels and found some bright points and filaments.

\section{Data Analysis} \label{sec:data}
\subsection{Nonlinear Force-free Field Modeling} \label{subsec:nlfff}
As mentioned in Section \ref{sec:intro}, coronal magnetic fields play an important role in flares. However, reliable high-accuracy magnetic field measurements are unavailable for the corona because the coronal lines have low brightness, are optically thin, and suffer from a large Doppler broadening. Currently, direct measurement of magnetic fields is only possible in low atmospheres close to the solar surface. The most effective method for inferring the magnetic field configuration in the corona is to extrapolate the magnetic field lines from the measured photospheric magnetic field. Thus, we performed a nonlinear force-free field (NLFFF) extrapolation using the MHD relaxation method, which 
solves the MHD equations directly in a zero-$\beta$ MHD approximation \citep{2014ApJ...780..101I}.
As the bottom boundary for the calculations, we used pre-flare magnetograms of the HMI obtained less than 1 h before the onset of the flares (Figures \ref{fig:101} (a) and (b)). We used a 472 $\times$ 472 grid maps available from SHARPs \citep{2014SoPh..289.3549B}. We changed the coordinates from the line of sight to Cartesian coordinates, where the x- and y-axes were defined on the surface and the z-axis was perpendicular to the surface. 
To satisfy a force-free boundary condition, we preprocessed the magnetograms by smoothing the $B_{x}$ and $B_{y}$ components to minimize the net Lorentz force and torque within 
the measurement errors \citep{2006SoPh..233..215W}. The magnetic density was normalized using a maximum absolute value of 4507 G for the confined flare and 4173 G for the eruptive flare. Thereafter, a 2 $\times$ 2 binning of the original magnetograms was performed (a pixel size was changed to 1.0 arcsecond) to flatten the pixels with a large deviation from the force-free state (Figures \ref{fig:101} (a) and (b)). 
Without preprocessing and binning, the NLFFF calculation was hard to converge because small scale magnetic field structures harm the force-freeness.
To initialize the NLFFF extrapolation, we calculated the potential field 
\citep{1982SoPh...76..301S} using the $B_{z}$ component of the magnetograms as the initial state in the NLFFF calculation. 
Further information regarding the NLFFF code used in this study is available in \citet{2014ApJ...780..101I, 2018ApJ...867...83I}. We set the grid size to $236 \times 236 \times 200$ (a grid equal to approximately 0.7 Mm), and iterated 10,000 times.

\subsection{Tracing the Magnetic Field Lines Related to the Energy Release} \label{subsec:ribbon}
In this study, we investigated instability-related parameters of the magnetic field lines chosen associated with bright flare kernels appeared in the first 1600 \AA\ frame immediately after the onset of the flares and derived similarity and differences observed between the confined and eruptive flares occurring in the same region on 2017 September 6. The reason why we focused only on the field lines originated from the kernels on the first frame is that they are highly involved in the initial trigger of energy release in the flares and thus we have imagined that instability-related parameters may show remarkable difference between the different types of flares. If we investigate the temporal evolution of the parameters for the magnetic field lines associated with the temporal changes of the flare kernels, we may derive new insights into the evolution of instabilities, which we leave for studies in the future. 

We defined the flare kernels as the region over 2000 DN/s/pixel (+18 $\sigma$ in the confined flare and +14 $\sigma$ in the eruptive flare calculated from $236 \times 236$ grids in the area of Figures \ref{fig:101} (a) and (b)) and recorded the coordinates of the flare kernels (the position of the flare kernels are shown as green areas in Figures \ref{fig:101} (c) and (d)). Note that the images were normalized to the exposure duration, and 1-$\sigma$ spread of the AIA 1600 \AA\ data is $53 \pm 108$ DN/s/pixel for the confined flare and $57 \pm 142$ DN/s/pixel for the eruptive flare.
We defined the bottom boundary at $z=0$ for NLFFF modeling. Flare kernels appear on the chromosphere at a height of a few megamerers; however, the NLFFF cannot accurately reflect the chromospheric influence owing to the difference in the force-free condition and a large grid size (approximately 0.7 Mm per a grid). \citet{2020ApJ...898...32K} debated the chromospheric influence on the NLFFF. 
We estimated that the accuracy of the co-alignment between the HMI and AIA was approximately one arcsecond by comparing the appearance of the sunspots in the HMI continuum data and the dark areas above the sunspots in the AIA 1600 \AA\ data. 
To evaluate the influence of the co-alignment uncertainty, we derived the standard deviation using the NLFFF data for the pixels surrounding the field line in Section \ref{subsec:result3}. In this study, we did not consider other causes of errors.

Investigating the average value and whole distribution of the twist number and decay index has already been shown by several authors, such as \citet{2017ApJL...849...L21}, \citet{2018ApJ...856...79Y}, \citet{2018A&A...619..A100}, \citet{2018ApJ...867L...5L}, and \citet{2019ApJ...870...97Z, 2020ApJ...890...10Z} for the flares (discussed in Section \ref{sec:discussion}). We considered that investigating the core of the flare energy release derived new insights into the flare eruption. Thus, using the 3D magnetic field derived from the NLFFF modeling, we traced the magnetic field lines from the flare kernels (the magnetic field lines are blue, red, orange, and green lines in Figures \ref{fig:101} (e) and (f), traced from each grid marked by green areas in Figures \ref{fig:101} (c) and (d)) and calculated the twist number, the largest value of the decay index and height (Z-coordinate) of each magnetic field line.
The twist number was calculated along each magnetic field line until $z=0$, and was recorded at the starting point. The decay index was calculated at the point where $\bm{B_{ex}}$ was calculated, using the horizontal component of the potential field. We chased along each magnetic field line and saved the largest decay index values because unlike MFRs, the magnetic field lines involved in the energy release have difficulty determining the apex position. The NLFFF was not used in the calculation of the decay index because it is a parameter for describing the background magnetic field, which is separate from the magnetic field induced by the current inside the flux rope. However, it is difficult for the NLFFF to distinguish between the background field constraining the flux rope and the induced field created by the flux rope \citep{2011ApJ...732...87C}. \citet{2012ApJ...748L...6N} compared the temporal and spatial decay index distribution calculated from the potential field and NLFFF, and found that calculating the potential field did not make much difference from what from NLFFF calculations. Thus, we chose the simpler MHD configuration, the potential field, to derive the decay index.
Moreover, the usage of the potential field for the decay index allows to compare our result more directly with former studies in which the potential field was used in the calculation of decay index—for example, \citet{2008ApJ...679L.151L}, \citet{2010ApJ...725L..38G}, \citet{2018ApJ...864..138J}, \citet{2018ApJ...867L...5L}, and \citet{2019ApJ...870...97Z, 2020ApJ...890...10Z}.
We also chased each magnetic field line to save the largest height values. Note that the relationship, 1 arcsecond = 0.731 Mm, which was determined from the distance between the Sun and Earth on 2017 September 6.

\section{Results} \label{sec:results}
\subsection{Spatial Distribution of the Magnetic Field Lines Involved in the Initial Flare's Energy Release} \label{subsec:result1}
Figures \ref{fig:101} (c) and (d) shows the first frame of AIA 1600 \AA\ measurements for (c) the confined and (d) eruptive flares, overlaid on the magnetograms in Figures \ref{fig:101} (a) and (b) at the photospheric level.
The frames show the areas in which the magnetic field lines involved in the initial energy release are rooted. In the confined flare, we observed three bright flare kernels that evolved into flare.
Two were on the positive and negative polarities north of the central positive polarity and the other was on the central positive and negative polarities near the center of the PIL. Four flare kernels were observed in the eruptive flare: one was the positive and negative polarities north of the central PIL, another on the central positive polarity near the center of the PIL, and others on the positive polarity south of the central positive polarity.

Figures \ref{fig:101} (e) and (f) shows the magnetic field lines extending from the pixels in the green areas shown in Figures \ref{fig:101} (c) and (d).
Thirty-one magnetic field lines were selected for the confined flare (blue and red lines in panel (e)), whereas 24 magnetic field lines (orange and green lines in panel (d)) for the eruptive flare. The seven red and green lines are the magnetic field lines with twist numbers greater than 0.3 in the confined flare, and greater than 0.6 in the eruptive flare (discussed in Sections \ref{subsec:result3}, and \ref{sec:discussion} in detail). The 24 blue and 17 orange lines are the other magnetic field lines in the confined and eruptive flares, respectively. Note that eight magnetic field lines in the eruptive flare with heights exceeding 100 Mm (approximately +1 $\sigma$ of all 32 magnetic field lines in the eruptive flare) are not shown because of the large uncertainties. Globally, the distributions of the magnetic field lines were similar in both cases. We classified the magnetic field lines into three groups: long sheared, connected to the northwest negative polarities; sheared, lying along the PIL at the center; and torus-shaped, appearing along the PIL at the southern end. Highly twisted seven magnetic field lines existed along the PIL at the centers of both flares.

\subsection{Properties of the Magnetic Field Lines Related to the Energy Release at the Initial Phase} \label{subsec:result2}
\begin{figure*}[ht!]
\centering
\includegraphics[scale=0.55,clip]{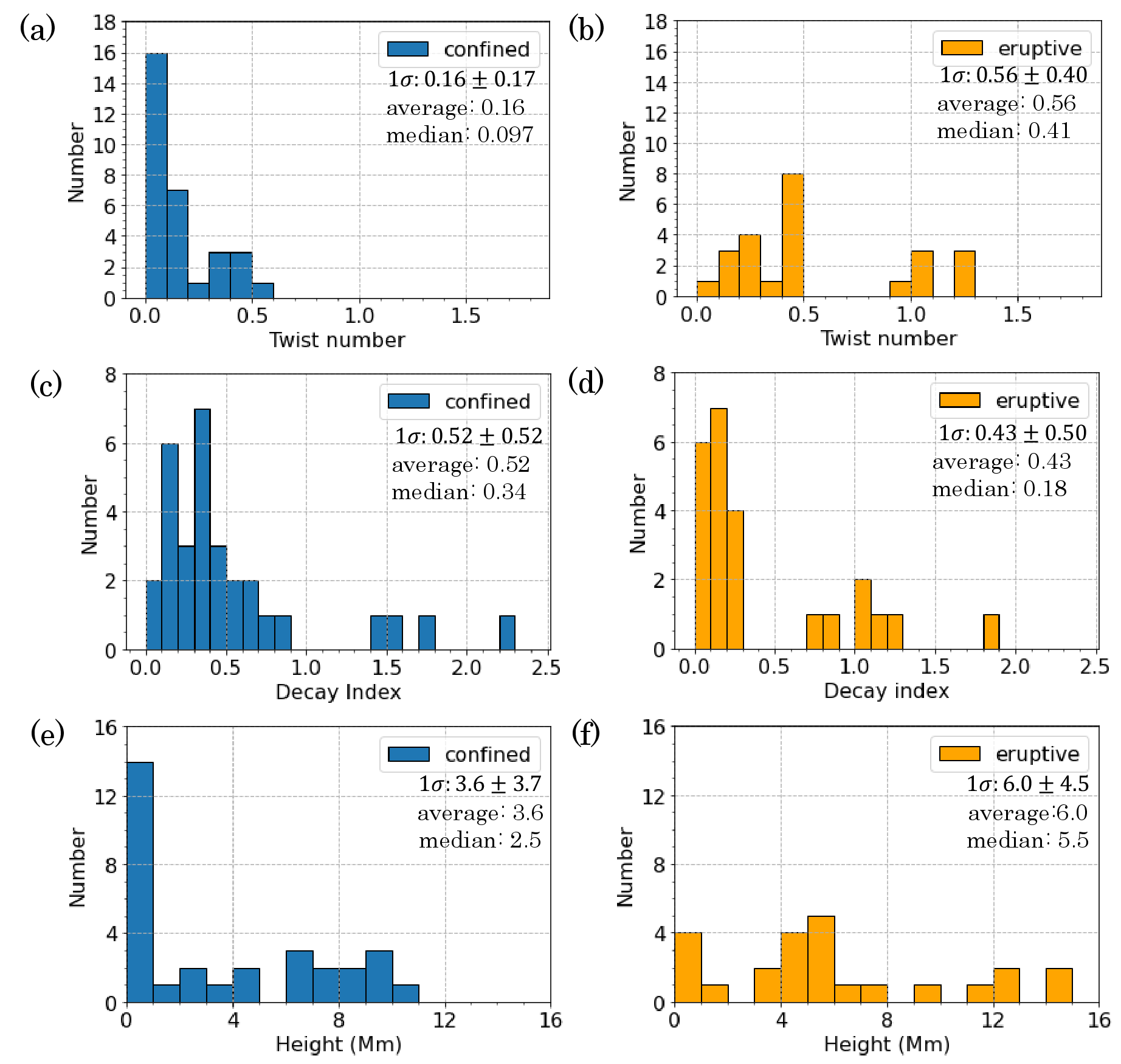}
\caption{Histograms of (a) and (b): twist number; (c) and (d): decay index; and (e) and (f): height (Mm) derived from each magnetic field line shown in Figures \ref{fig:101} (e) and (f). 
The left side shows the confined flare, and the right side shows the eruptive flare. The vertical axes represent the number of magnetic field lines. On the upper right side of each histogram, the 1-$\sigma$ spread, average value, and median value are presented.
\label{fig:39}}
\end{figure*}

Figure \ref{fig:39} shows histograms of the twist number, decay index, and maximum height (for quantitatively checking the geometrical differences) measured for the energy-release related lines at the start of the confined and eruptive flares. 
The confined flare is shown on the left, and the eruptive flare is shown on the right. Figures \ref{fig:39} (a) and (b) show that the eruptive flare had a larger twist number in seven lines, which was not observed in the confined flare. The confined flare was distributed between 0–0.6 in succession, whereas the eruptive flare was distributed between 0–0.5 and 0.9–1.3. There was a gap in the twist number between 0.5–0.9, and most magnetic field lines existed below 0.5. The minimum and maximum values for the confined flare were 0 and 
0.58, respectively, while those for the eruptive flare were 0.02 and 1.29, respectively. In the confined flare, there were no twist numbers above 0.6. The average value for the confined flare was 0.16, whereas that for the eruptive flare was 0.56. The average value in the eruptive flare was 3.5 times larger. The median value in the confined flare was 0.097 and that in the eruptive flare was 0.41. The 1-$\mathrm{\sigma}$ spread in the confined flare was $\rm{0.16 \pm 0.17}$, and that in the eruptive flare was $\mathrm{0.56 \pm 0.40}$.

Figures \ref{fig:39} (c) and (d) show that the decay index distribution did not exhibit clear differences between the confined and eruptive flares. The number distribution of the decay index showed two peaks in both flares: the confined flare distributions were 0–0.9 and 1.4–2.3, and the eruptive flare distributions were 0–0.3 and 0.7–1.9. The minimum and maximum values in the confined flare were 0.05 and 2.23, respectively. The values for the eruptive flare 
were 0 and 1.88, respectively. The average value for the confined flare was 0.52, whereas that for the eruptive flare was 0.43. The median value for the confined flare was 0.34, whereas that for the eruptive flare was 0.18. The 1-$\mathrm{\sigma}$ spread in the confined flare was 
$\mathrm{0.52 \pm 0.52}$ and that in the eruptive flare was $\mathrm{0.43 \pm 0.18}$.

Figures \ref{fig:39} (e) and (f) show that most of the magnetic field lines in the confined flare were concentrated at the lowest height, and five lines in the eruptive flare had height greater than 11 Mm, which were not observed in the confined flare. The confined flare had one peak at 0–1 Mm with a broad distribution up to 11 Mm, whereas the eruptive flare had a scattered distribution with three groups at 0–2, 3–8, and 9–15 Mm. The minimum and maximum values in the confined flare were 0.0124 and 10.3 Mm, respectively, and those in the eruptive flare were 0.00073 and 14.8 Mm, respectively. 
The average value in the confined flare is 3.6 Mm and that in the eruptive flare is 6.0 Mm. The median 
value in the confined flare was 2.5 Mm and that in the eruptive flare was 5.5 Mm. The 1–$\mathrm{\sigma}$ spread in the confined flare was $\mathrm{3.6 \pm 3.7}$ Mm and that in the eruptive flare was $\mathrm{6.0 \pm 4.5}$ Mm.

\subsection{Characteristics of the Seven Highly Twisted Magnetic Field Lines} \label{subsec:result3}
Section \ref{subsec:result2} shows that the twist number differs between the confined and eruptive flares, and the characteristics of the seven magnetic field lines that have larger twist number values in the eruptive flare were investigated. 
To examine the characteristics of the highly twisted lines from the perspective of torus instability, a scatter plot of the twist number versus the decay index was created. A scatter plot of the twist number versus height was added as a property of the magnetic field line geometry, because reconnection at a higher point may be important for generating the eruptive flare. It is difficult to estimate all uncertainties in the NLFFF model because there are many problems associated with the uniqueness of NLFFF model solutions. Note that \citet{2012LRSP....9....5W} reviewed these problems. However, it is worth understanding how the derived magnetic field lines may change in association with the surrounding magnetic field lines. We used nine magnetic field lines from the pixels surrounding the field line of interest and derived three physical parameters, providing their standard deviation as a proxy to show how much they may change depending on the selection of pixels in the flare kernels and the co-alignment between the HMI and AIA data. The magnetic field lines in which the twist number, decay index, or height were outside the 3-$\sigma$ spread were eliminated before calculating the error.

\begin{figure*}[ht!]
\plotone{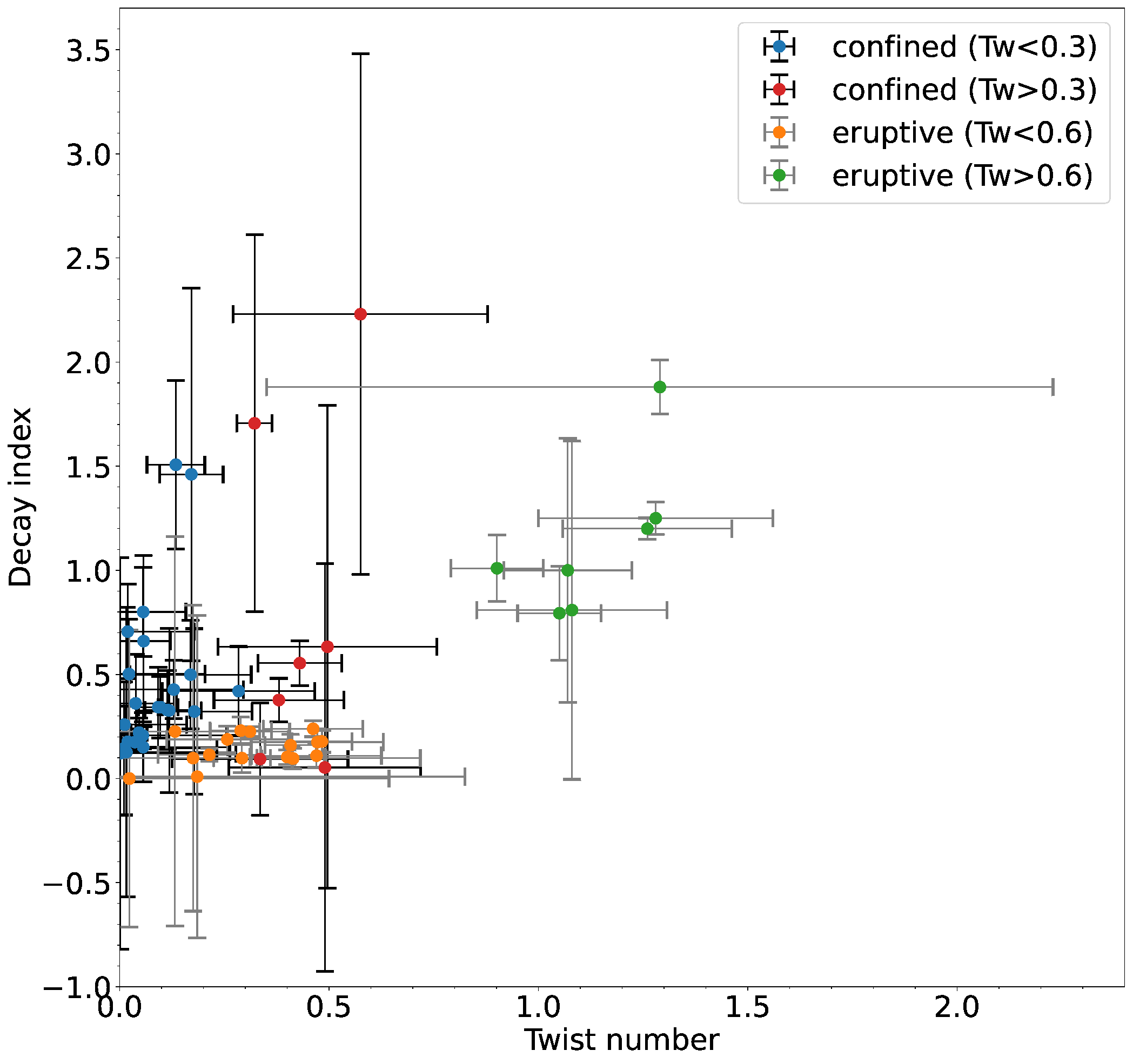}
\caption{Scatter plot between the twist number (horizontal axis) and decay index (vertical axis) derived from each magnetic field line.
The blue dots are the magnetic field lines that have a twist number below 0.3 for the confined flare, red dots are above 0.3 for the confined flare, orange dots are below 0.6 for the eruptive flare, and green dots are above 0.6 for the eruptive flare.  
\label{fig:41}}
\end{figure*}

Figure \ref{fig:41} shows a scatter plot of the twist number and decay index. The error bars show the standard deviation derived using nine magnetic field lines extending from the pixels surrounding the magnetic field line of interest. Note that the extremely deviated data, i.e., 3-$\sigma$ or larger deviation, was excluded to calculate the standard deviation for each footpoint.
The highly twisted magnetic field lines were separated from the less twisted lines only in the eruptive flare. Considering the error, the twist number primarily causes the difference and not the decay index. The magnetic field lines were concentrated in the lower-left region for both flares. Including the error bars, the magnetic field lines in the confined flare existed for twist numbers below 0.9. Although the confined flare cannot be separated, the eruptive flare can be completely separated between the highly twisted ($T_{w} > 0.9$) seven magnetic field lines (green points) and the less twisted ($T_{w} < 0.6$) 17 magnetic field lines (orange points). In the confined flare, the decay index distribution was similar for the highly and lowly twisted lines. In the eruptive flare, highly twisted lines may have larger decay index values; however, this tendency was not clearly identified in the error range.

\begin{figure*}[ht!]
\plotone{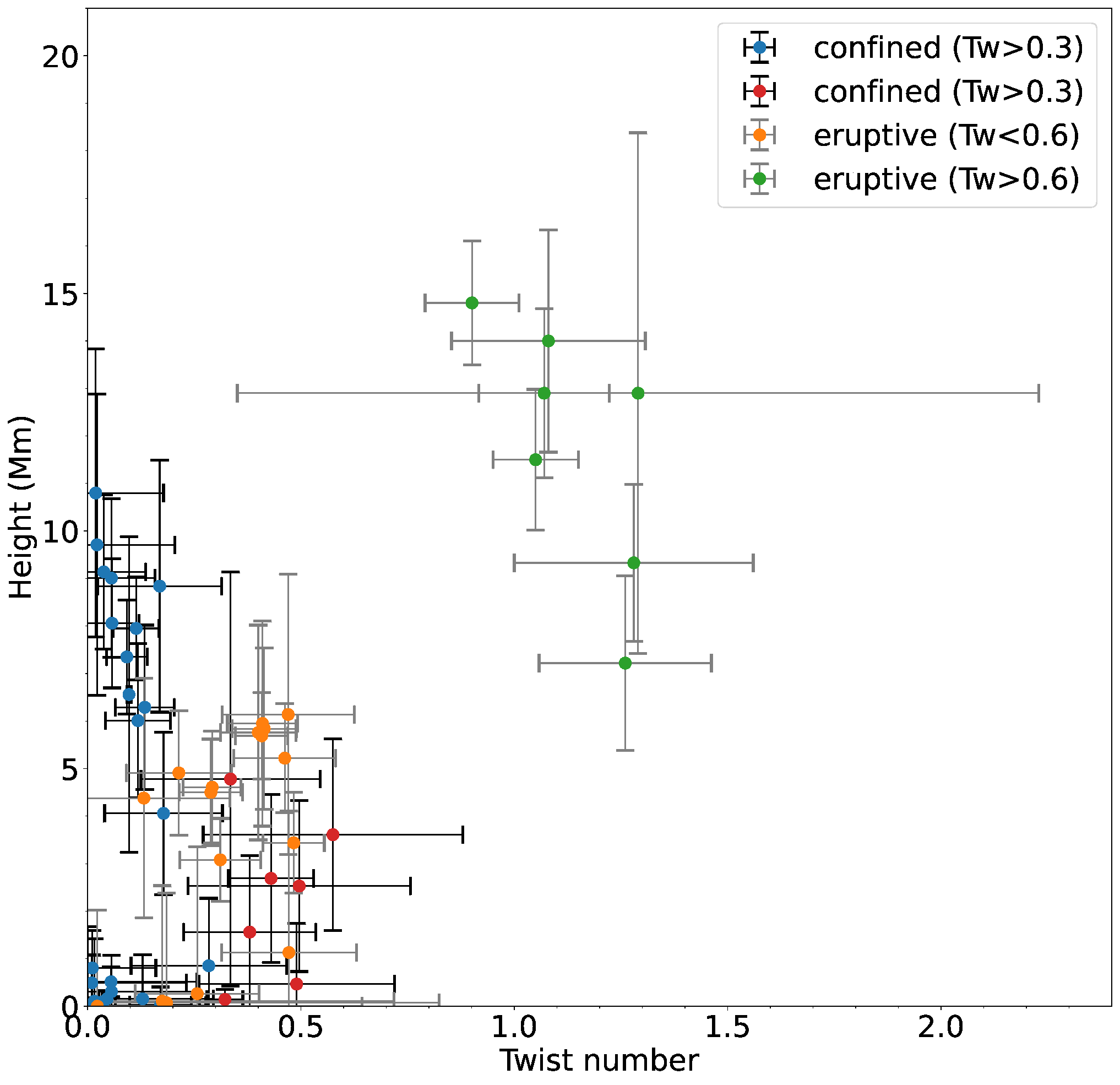}
\caption{Scatter plot between the twist number (horizontal axis) and height (vertical axis) derived from each magnetic field line. The symbols and colors are used in the same way as in Figure \ref{fig:41}.
\label{fig:42}}
\end{figure*}

Figure \ref{fig:42} is a scatter plot of the twist number and height.
The magnetic field lines were concentrated in the lower-left region in both flares, except for seven highly twisted lines in the eruptive flare.
The eruptive flare could be completely separated between the highly twisted ($T_{w} > 0.6$) seven magnetic field lines (green points) and the less twisted ($T_{w} < 0.6$) 17 magnetic field lines (orange points); all lines in the confined flare were excluded. In the confined flare, we divided the data points into two groups: the lower-left side, where the height is below 2 Mm, and the upper right side, where the height decreases with increasing twist number (seven highly twisted lines have lower height). No magnetic field lines existed in the region where the twist number was greater than 0.9 and the height was less than 5 Mm.

\section{Discussion} \label{sec:discussion}
First, we discuss the relationship between each parameter and the eruptivity. 
Figures \ref{fig:39} (a) and (b) indicate that most magnetic field lines involved in the initial energy release of the flares exist in a similar range of twist numbers; however, some magnetic field lines in the eruptive flare have a larger twist number than all lines in the confined flare. Kink instability may indicate differences in eruptivity. Comparing the soft X-ray flux with the average twist number, the difference in the flare magnitude may arise primarily from the magnetic free energy of the magnetic field lines related to the initial flare energy release. \citet{2017ApJL...849...L21} studied the eruptive flare and derived an average twist number of approximately 1.5 for the inner portion of the MFR and approximately 1.2 for the outer portion. \citet{2018ApJ...856...79Y} derived an average twist number of 0.55 for the entire MFR in the confined flare and 0.7 for the eruptive flare. \citet{2018A&A...619..A100} found that the MFR axis in the eruptive flare exceeds 1.75. \citet{2018ApJ...867L...5L} derived slightly larger values in the average twist number for both flares (1.3–1.4 for the confined and 1.3–1.5 for the eruptive flares). These studies did not derive clear differences in the twist number between the confined and eruptive flares; however, our study shows that a limited number of magnetic field lines connected to the flare ribbons in the initial triggering phase may exhibit different twist number magnitudes. The study result does not reach the theoretical kink instability threshold of 1.75; however, this may be because the magnetic field lines involved in the flare energy release may change the twist number abruptly by magnetic reconnection. An interpretation of Figures \ref{fig:39} (c) and (d) is that torus instability may not have been the primary factor for the eruption, and the background external poloidal magnetic flux density condition was similar in both the flares. \citet{2018ApJ...867L...5L} derived the average decay index value above the flaring PIL with height and showed that the average height where the decay index became 1.5 (a theoretical torus instability threshold) is the same in the flares. \citet{2019ApJ...870...97Z, 2020ApJ...890...10Z} made the contour of the theoretical threshold value ($n = 1.5$) to be the torus unstable state, and showed that the decay index distribution does not change; however only the MFR (defined by the area of the twist number over 1) reaches a decay index greater than 1.5 in the eruptive flare because the MFR expands. 
Figures \ref{fig:39} (e) and (f) show that the height distribution is similar between the confined and eruptive flares; however, many lines concentrate at 0–1 Mm only in the confined flare, and several lines exceed 11 Mm in height only in the eruptive flare. This might indicate that the initial reconnection height of some lines affects eruptivity.

Next, we discuss the features of the highly twisted magnetic field lines, which may characterize eruptivity, by examining the scatter plot of the twist number versus the decay index or height. Figure \ref{fig:41} shows that torus instability seems to have less influence on eruptivity than kink instability, and there is no difference between the confined and eruptive flares in the majority of the magnetic field lines involved in the initial triggering process. However, only a limited number of lines in the eruptive flare influence eruptive behavior. Comparing the scatter plots of \citet{2015Natur.528..526M} and \citet{2018ApJ...864..138J}, Figure \ref{fig:41} shows five signatures: the magnetic field lines in both flares concentrate on the lower twist number and decay index side (stable regime); a few lines in the confined flare exist in the lower twist number and higher decay index side (failed torus regime); 
the decay index does not have a threshold to be the eruptive flare; the threshold of the twist number is 0.9; and three lines in the eruptive flare exist over the threshold. 

Figure \ref{fig:42} indicates that some magnetic reconnection in the eruptive flare may have occurred with both larger magnetic free energy and higher altitude. The majority of the magnetic field lines of both flares were concentrated on the lower-left side, where the twist number and height were lower and with similar distributions, indicating that the dominant reconnection takes place at similar heights. The taller lines in the confined flare were potential-like long lines, whereas those in the eruptive flare were highly twisted, perhaps because they were longer and easier to twist. According to \citet{2020Sci...369..587K}, the flare center is located close to the north of the central PIL in the confined flare, and around the center of the central PIL in the eruptive flare. Figures \ref{fig:101} (e) and (f) shows that less twisted but taller lines primarily exist north of the central PIL in the confined flare and are highly twisted, whereas higher lines exist in the center of the central PIL in the eruptive flare. Thus, the flare center moves from the area where the altitude is higher but lowly twisted lines exist, to the area where the seven highly twisted and taller lines exist.

We speculate how the magnetic configurations of these
flares can be summarized by these field lines as follows.
Both flares occurred in a similar magnetic distribution, in which most of the magnetic field lines were under similar MHD conditions. The magnetic field lines at the south in the eruptive flare (the green lines in Figure \ref{fig:101}), however, have a remarkable feature, that is, highly twisted and tall. A small number of the field lines may initially reconnect in more kink unstable state or magnetic free energy, and at higher point. This may mean that the reconnection is more intense and closer to space, and stimulate flare eruption.

\section{Conclusions} \label{sec:conclusions}
An observational understanding of MHD instabilities can help describe the onset process and variety of solar flares, which would facilitate geo-effective flare predictions. In this study, we compared the confined and eruptive flares that occurred at the same location and only 3 h apart on 2017 September 6, using the twist number, maximum decay index, and maximum height in a similar MFR configuration. The three average parameters of the MFR and each magnetic field line involved in the initial flare energy release were investigated. We identified these lines from the flare kernel positions and extrapolated the coronal magnetic field lines using NLFFF modeling with photospheric vector magnetograms. We found that: 1) The position of the initial flare kernels and the configuration of the magnetic field lines were similar between the confined and eruptive flares. 2) The twist numbers of all the magnetic field lines in the confined flare were below 0.6, whereas the eruptive flare had magnetic field lines with twist numbers up to 1.3. However, the decay index distributions did not exhibit any clear differences. 3) Although most magnetic field lines have similar values for the three parameters between the confined and eruptive flares, seven out of twenty-four magnetic field lines in the eruptive flare have larger twist number and height values. Our study suggests that part of the magnetic field lines in the kink instability state may be important for eruptive flares, and a higher altitude may make it easier for flares to erupt in terms of the altitude of magnetic reconnection. Further investigations will be valuable in the temporal evolution of the parameters in association with the temporal changes of flare kernels to derive new insights into the evolution of instabilities, which we leave for studies in the future.

\begin{acknowledgments}
We gratefully acknowledge the \textit{SDO}/HMI and \textit{SDO}/AIA teams for providing data. Our NLFFF calculations were performed using the JAXA Supercomputer System Generation 3 (JSS3). We thank Dr. Satoshi Inoue at the Center for Solar-Terrestrial Research, New Jersey Institute of Technology, for providing the NLFFF code on JSS3 and the preprocessed magnetograms. 
We thank the anonymous referee for the careful and valuable comments.
TS’s contribution to this study is partially supported by JSPS KAKENHI grant No. 18H05234 and the grant of OML Project by the National Institutes of Natural Sciences (NINS program No. OML032402).
\end{acknowledgments}

\bibliography{sample631}{}
\bibliographystyle{aasjournal}

\end{document}